\documentclass[amsmath,amssymb,superscriptaddress,doublecol]{epl2}
\usepackage{amsmath,amssymb,bbm}
\usepackage{graphicx} 
\usepackage{xcolor}
\usepackage{float}

\begin{document}

\title{Localization versus subradiance in three-dimensional scattering of light}%

\author{N. A. Moreira\inst{1,2} \and R. Kaiser\inst{2} \and R. Bachelard\inst{2,3}}

\institute{

  \inst{^1} Instituto de F\'{i}sica de S\~{a}o Carlos, Universidade de S\~{a}o Paulo - 13560-970 S\~{a}o Carlos, SP, Brazil\\
  \inst{^2} Universit\'{e} C\^{o}te d'Azur, CNRS, INPHYNI, France \\
  \inst{^3} Departamento de F\'{\i}sica, Universidade Federal de S\~{a}o Carlos, Rod. Washington Lu\'{\i}s, km 235 - SP-310, 13565-905 S\~{a}o Carlos, SP, Brazil
}
\date{\today}


\abstract{
We study the scattering modes of light in a three-dimensional disordered medium, in the scalar approximation and above the critical density for Anderson localization. Localized modes represent a minority of the  total number of modes, even well above the threshold density, whereas spatially extended subradiant modes predominate. For specific energy ranges however, almost all modes are localized, yet adjusting accordingly the probe frequency does not allow to address these only in the regime accessible numerically. Finally, their lifetime is observed to be dominated by finite-size effects, and more specifically by the ratio of the
localization length to their distance to the system boundaries.
}

\maketitle

\section{Introduction}
Since the pioneering work by Philip W. Anderson to explain the transition from metal to insulator for electronic transport~\cite{AndersonPaper}, trapping of waves in a disordered potential has been reported in a variety of systems, ranging from acoustics to matter waves~\cite{Hu2008,Billy2008,Kondov2011,Semeghini2015,Chabanov2000}. However, for electromagnetic waves, even though their localization has been reported in 1D~\cite{Berry1997} and 2D~\cite{Laurent2007}, the observation of Anderson localization in 3D has been the subject of an ongoing controversy~\cite{Skipetrov2016}. Indeed, the initial signatures used in the diffuse transmission or in the late time dynamics were later questioned~\cite{Genack1991,Wiersma1997,Schuurmans1999,Sperling2012,Strzer2006,Scheffold1999,Sperling2016}, and at the moment, an unambiguous experimental signature is lacking.

From a fundamental point of view, the vectorial nature of electromagnetic waves has been shown to prevent localization~\cite{SkipetrovAbsence,Maximo,Bellando2014}. This theoretical analysis was based on the scaling analysis~\cite{Abrahams1979}, where the linewidth of the modes (i.e., their inverse lifetime) is compared to the typical difference of frequency between them. The transition to localization is thus marked by the emergence of modes with long lifetimes.

On the other side, collective scattering modes can have diverse origins. A paradigmatic example was introduced with superradiance from a fully inverted system of many two-level systems. R. Dicke showed that this can lead to an accelerated emission by cascading between symmetric collective atomic states~\cite{Dicke1954}, a prediction that was later confirmed experimentally~\cite{Skribanowitz1973}.
Another intriguing collective scattering effect from a single excitation among many two-level scatterers has been reported in Ref.~\cite{Guerin2016}, where subradiance with radiation rates below the single scatterer rate has been observed.
However, because they can be observed in dilute regime, i.e., well below the critical density for localization, these long lifetimes are unrelated to localization.

The presence of subradiance in the dilute limit clearly questions the ability of late-time dynamics to capture the localization transition~\cite{Strzer2006}. But more generally, it highlights the necessity to distinguish carefully different kinds of modes, in order to be able to address specifically the localized ones. In particular, superradiance and subradiance may provide competing dynamical signatures to localization, and they exist even in absence of disorder~\cite{Cottier2018}. It is thus necessary to understand the interplay between these phenomena, in order to be able to determine an unambiguous temporal signature of the localization transition.

In this paper, we focus on a scalar model of light scattering in 3D. Even though it is known that the full vectorial model does not present a localization transition~\cite{SkipetrovAbsence}, the addition of a strong magnetic field was shown to split the scattering channels and reproduce the localization transition of the scalar model~\cite{Skipetrov2015,Cottier2019}. In this scalar model, the transition to localization  has been characterized using the eigenvalues and eigenvectors of the interaction matrix~\cite{SkipetrovAbsence,Bellando2014,Skipetrov2018,Skipetrov2018b}.

In this work, we study the competition between localized, subradiant and superradiant modes, showing that a transition to the localization regime is not achieved for the system sizes accessible numerically. Indeed, localized modes represent only a minority of modes, and despite their high densities in a specific range of energies, even an appropriate choice of pump frequency does not prevent from strongly coupling to the other modes. Finally, the lifetime of these modes appear to be driven by a leakage due to the finite distance to the boundaries of the system. This allows us to evaluate the ratio of the localization length to their distance to the system's boundaries from the lifetime of the localized modes. These results contribute to the difficult task of finding unambiguous signatures of Anderson localization of light in 3D.


\section{Coupled Dipole Model and Scattering Modes}\label{Sec:Model}
We consider a model of $N$ point-like scatterers with positions $\mathbf{r}_j$ distributed randomly and homogeneously inside a spherical cloud of radius $R$. The scatterers can be thought as two-level atoms with linewidth $\Gamma$ and transition frequency $\omega_a$. The system is illuminated by a monochromatic wave characterized by the Rabi frequency $\Omega(\mathbf{r})$ of atom-light interaction and frequency $\omega=kc$, detuned by $\Delta=\omega-\omega_a$ from the atomic transition. The exchange of virtual and real photons between the particles result in an effective interaction between the atomic dipoles $\beta_j$. In the linear optics regime, the dipole dynamics is governed by a set of coupled differential equations~\cite{Lehmberg1970}:
\begin{equation}
\label{eq1}
    \small \frac{d\beta_j}{dt} = \left ( i\Delta-\frac{\Gamma}{2} \right )\beta_j  - \frac{\Gamma}{2}\sum_{m\ne j} \frac{ \exp (i k  |\mathbf{r}_j-\mathbf{r}_m|) }{i k |\mathbf{r}_j-\mathbf{r}_m|}\beta_j - \frac{i\Omega(\mathbf{r}_j)}{2}.
\end{equation}
We note that the model can be derived from fully classical principles~\cite{Svidzinsky2010,Cottier2018}. The first right-hand term describes the single-dipole dynamics, the second one corresponds to the dipole-dipole coupling, and the last one to the driving of the incident pump field. In order to avoid contributions from pair physics (2 closeby atoms), we set an exclusion volume in the particle distribution, based upon the density: $r_\mathrm{min} = \rho^{-1/3}/\pi$~\cite{Michelle}. Throughout this work, the pump considered is a plane-wave travelling in the $\hat{z}$ direction: $\Omega(\mathbf{r}_j) = \Omega e^{i\mathbf{k}_0.\mathbf{r}_j}$, with $\mathbf{k}_0=k\hat{z}$; we have checked that using a Gaussian beam with a waist comparable to the system size does not alter our conclusions.

The dipole-dipole coupling results in collective scattering modes, with an associated decay rate and frequency. These correspond to the eigenvectors $\Psi^n$ of Eq.\eqref{eq1}, which conveniently written in matrix form:
\begin{equation}
\label{eq2}
  \frac{d\vec{\beta}}{dt}   = \left(i\Delta\mathbbm{1}_N+\mathbf{M}\right) \vec{\beta} -i \frac{\vec{\Omega}}{2},
\end{equation}
where $ M_{jm}=-\Gamma/2[\delta_{jm}+(1-\delta_{jm}) \exp(i k|\mathbf{r}_j-\mathbf{r}_m|)/(ik |\mathbf{r}_j-\mathbf{r}_m|)]$, with $\delta_{jm}$ the Kronecker symbol. The eigenvalues of $M$, $\lambda_n$ decompose as the decay rates $\Gamma_n = -\Re(\lambda_n)$ and frequency shifts $\omega_n = \Im(\lambda_n)$ from the atomic transition. The components $|\Psi^n_j|^2$ represents the excitation contribution of atom $j$ to mode $n$. 

\begin{figure}
   \centering
    \includegraphics[width=\columnwidth]{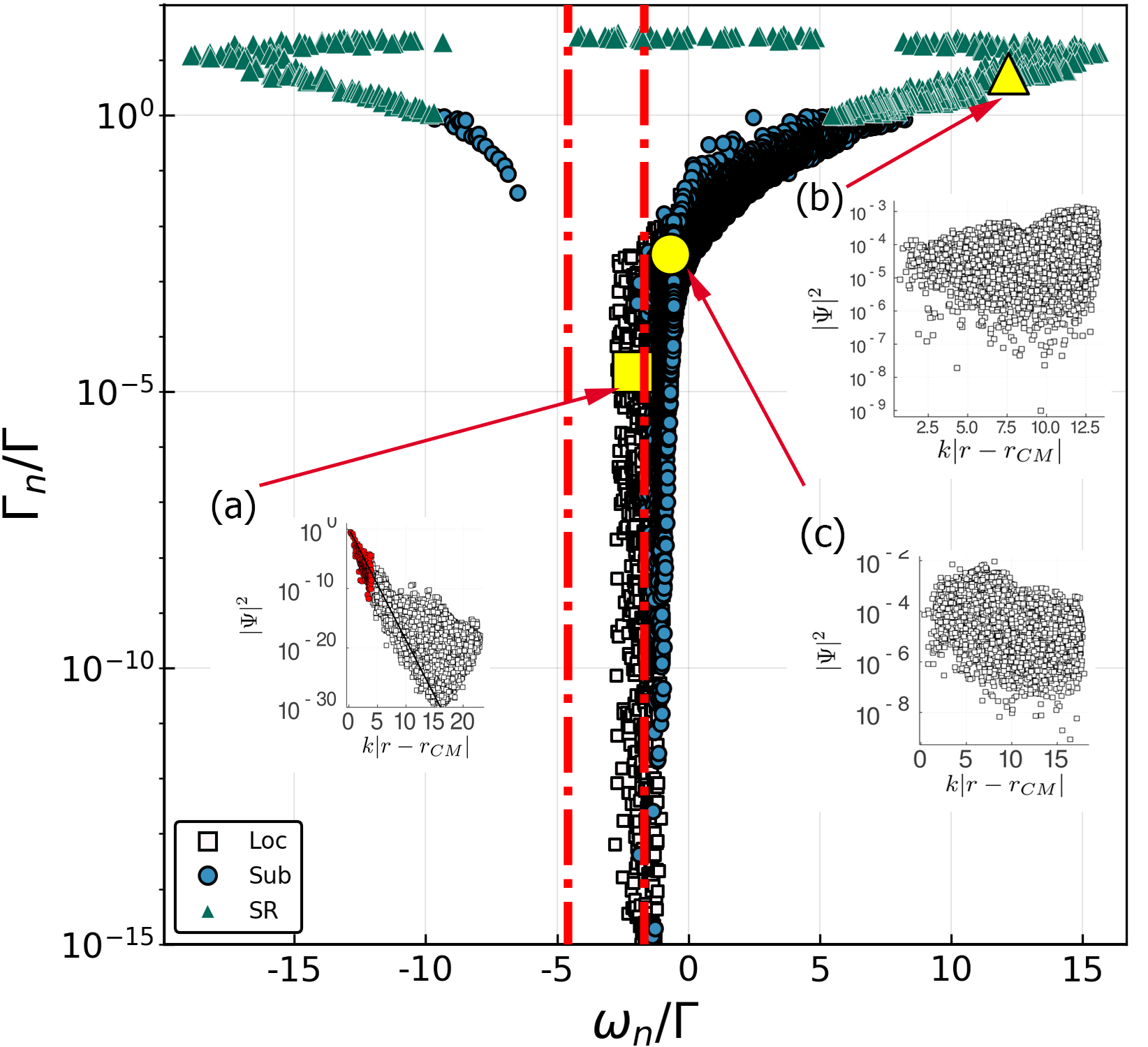}
    \caption{Eigenspectrum $(\Im(\lambda_n), \Re(\lambda_n)$ for a system with a density ($\rho=10\rho_c=k^3$) ten times larger than the critical density, and $N=5000$. Superradiant, subradiant and localized modes are indicated by triangles, circle and square markers, respectively. The insets show typical spatial profiles of each kind of mode. For the localized mode (a), the red markers denote the particles used to determine the localization length of the mode, and the line to the corresponding fit. The vertical lines correspond to the Ioffe-Regel prediction, including the Lorentz-Lorenz shift (see Eq.\eqref{eq:Ioffe}).}
    \label{Fig1}
\end{figure}

Superradiant and subradiant modes emerge from the dipole-dipole coupling, as well as exponentially localized modes above the transition threshold $\rho_c\approx 0.1k^3$~\cite{SkipetrovAbsence,Bellando2014}. We decompose modes into three categories, which are illustrated in Fig.\ref{Fig1}, along the eigenspectrum: (a) Localized modes, that present an exponentially decaying spatial profile [see Fig.\ref{Fig1}(a)]; (b) Superradiant modes, that present a decay rate $\Gamma_n>1$ [see Fig.\ref{Fig1}(b)]; (c) Subradiant modes ($\Gamma_n<1$) that do not present an exponentially decaying spatial profile [see Fig.\ref{Fig1}(c)].

The spatial profiles presented in the insets of Fig.\ref{Fig1} describe the particles contribution, ordered from their distance to the mode center-of-mass $\mathbf{r}_\mathrm{CM}^n=\sum_m |\Psi^n_m|^2 \mathbf{r}_m$. They are typical profiles of the three kinds of modes encountered in the system. Superradiant and subradiant are rather extended, whereas localized modes present, for short distances, an exponential decay.
Indeed, in the open system~\eqref{eq1}, localized modes present a spatial profile which decays exponentially at first, before presenting a much slower decay~\cite{Biella2013,Celardo2017}. The determination of the localization length $\xi$ that characterizes the decay ($\Psi(\mathbf{r})\sim \exp(|\mathbf{r}-\mathbf{r}_\mathrm{CM}|/\xi)$) of these hybrid modes must thus be realized by giving a larger weight to the atoms contributing most to the mode. This is achieved using the $L_1$ minimization fitting procedure to compute $\xi$ from the $|\Psi(\mathbf{r})|^2$ profile, without logarithmic rescaling, considering only the atoms closest to the mode center-of-mass, such that the remaining ones represent less than $10^{-4}$ of the mode norm. The minimization of the $L_1$ aims to minimize the \textit{absolute difference} of errors (quantified by the Adapted Coefficient of Determination $R^1$), rather than the usual square error (i.e., the $R^2$), and has been used to overcome outliers issues~\cite{l1minimization1, l1minimization2, l1minimization3}. Throughout this work, we require for a mode to present a $R^1$ larger than $0.6$ to be considered localized: An inspection of the profiles of various modes suggests that this value is a reasonable choice to distinguish the modes. 

\begin{figure}
   \centering
    \includegraphics[width=\columnwidth]{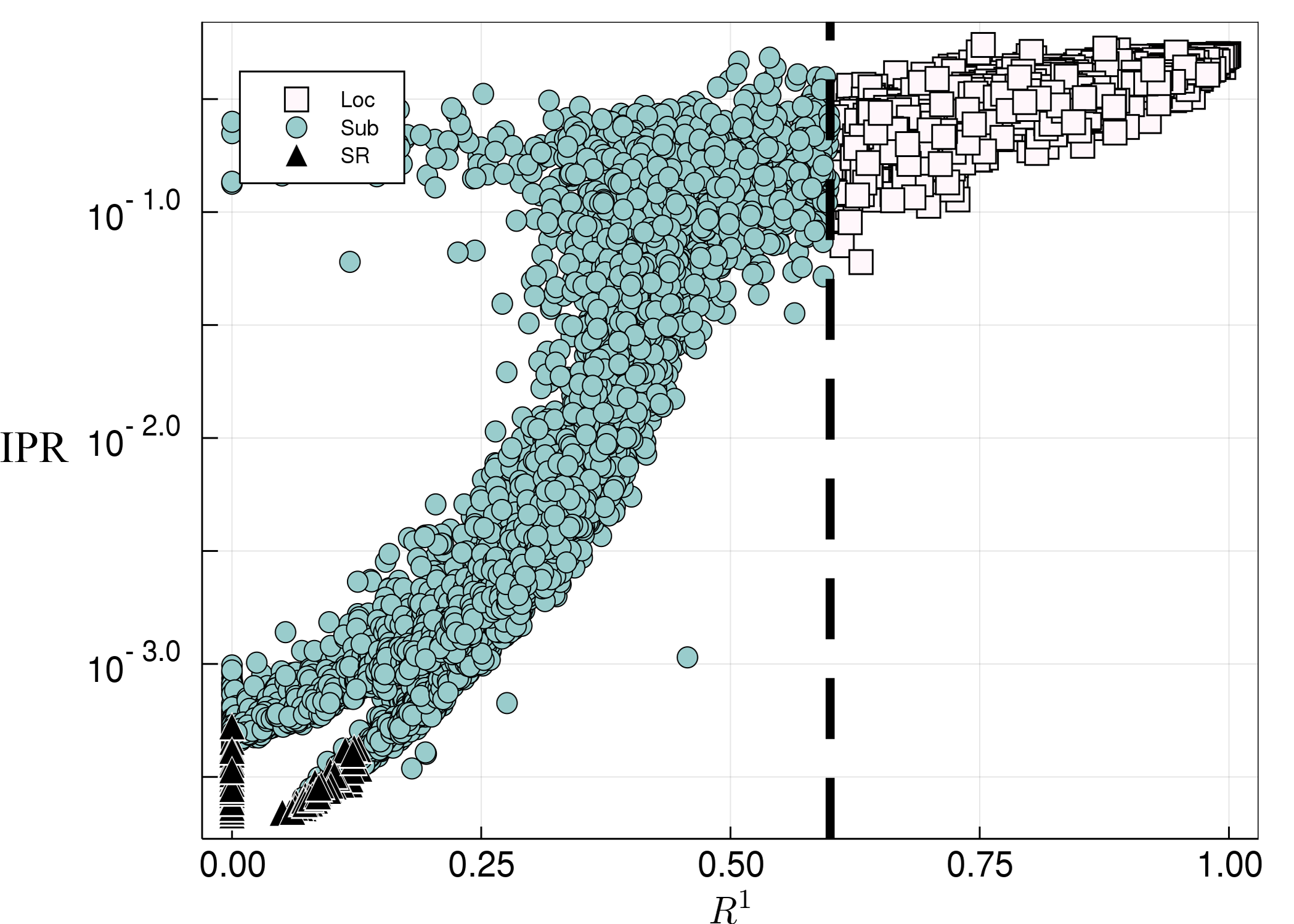}
    \caption{Relation between the IPR and $R_1$ of the scattering modes. The localized/subradiant/superradiant modes are marked as squares/disks/triangles. Same parameters as Figure~\ref{Fig1}.}
    \label{FigIPR}
\end{figure}

Superradiant and subradiant modes present a broad spatial profile whereas localized modes are characterized by the contribution of very few atoms (see Fig.~\ref{Fig1}(a) and (b)). This is confirmed by the analysis of the Inverse Participation Ratio (IPR), defined as
\begin{equation}
    \mathrm{IPR}_n=\frac{\sum_m |\Psi_m^n|^4}{\left(\sum_m |\Psi_m^n|^2\right)^2}.
\end{equation}
As can be observed in Fig.\ref{FigIPR}, superradiant modes present an IPR of $\sim 10^{-3}$ (the simulations were realized for $N=5.10^3$, which naturally bounds the IPR to a lower value of $1/N=1/(5.10^3)$), while for localized modes present it is typically $0.2-0.5$. Subradiant modes exhibit a much broader variety of IPR, from $\sim0.4$ to $10^{-3}$, that witness their extended nature. Thus, despite the IPR has been used extensively to study the Anderson localization transition~\cite{Bell1972}, the coexistence of exponentially localized and extended subradiant modes highlights the importance of monitoring directly the spatial profiles of the modes.
 





\section{Distribution of modes and their coupling to the exterior world}\label{Sec:Coupling}

\begin{figure}
   \centering
    \includegraphics[width=\columnwidth]{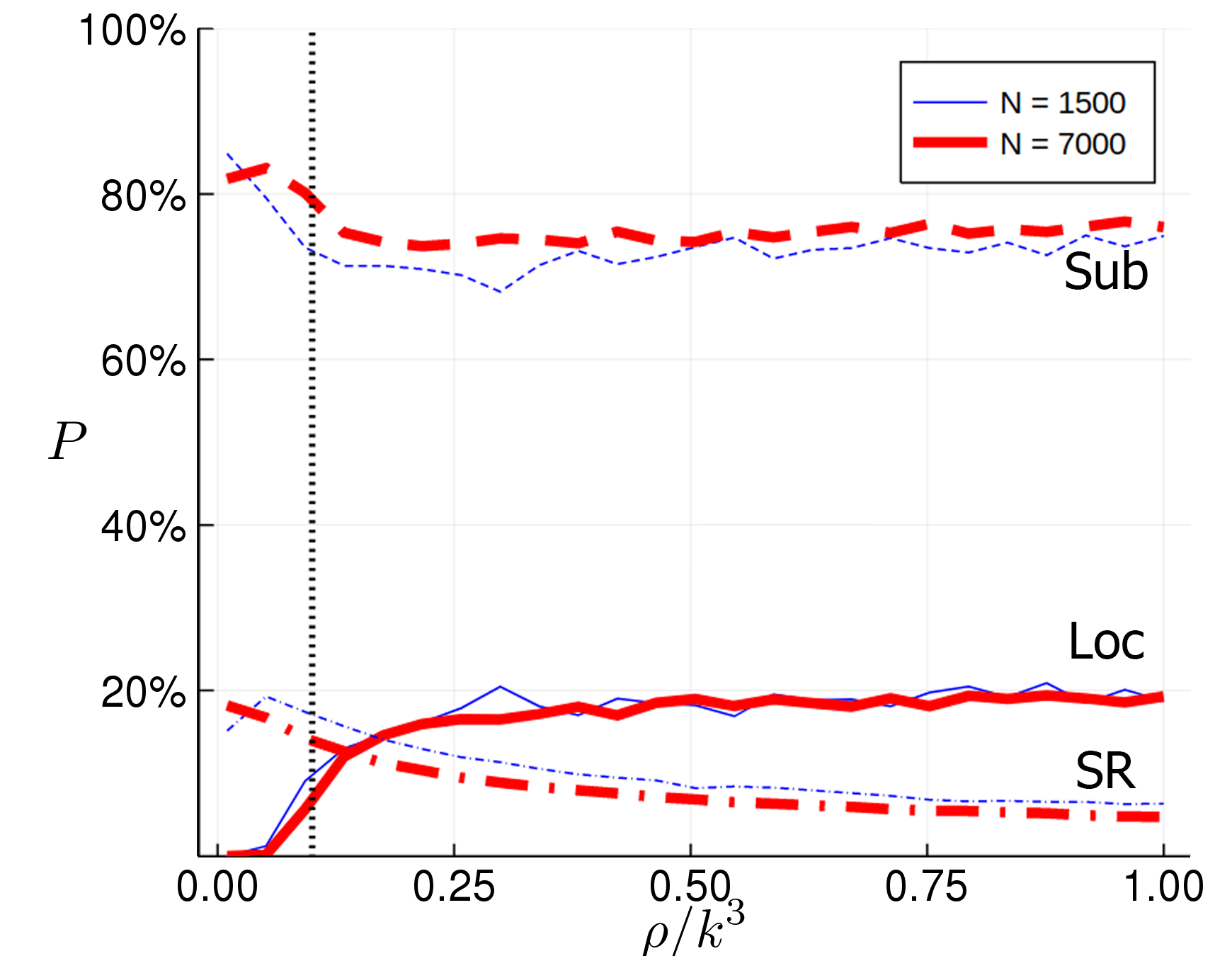}
    \caption{Proportion of localized (plain line), subradiant (dashed) and superradiant (dash-dotted) modes for an increasing density and a fixed particle numbers: $N=1500$ (thin curves) and $7000$ (thick curves).}
    \label{Fig2}
\end{figure}

Figure~\ref{Fig1} suggests that, even at a density ten times larger than the critical density for localization, a significant number of modes do not present an exponentially decaying profile: the scattering modes present a broad range of energies, the localized modes being concentrated over narrow band. 
Interestingly, the Ioffe-Regel criterion appears to predict correctly the edge between the localized and subradiant modes, once the Lorentz-Lorenz shift is included~\cite{Kaiser2000,Kaiser2009,Cottier2019}:
\begin{equation}
\frac{\Delta_c}{\Gamma}=-\frac{\pi\rho}{k^3}\pm \frac{1}{2}\sqrt{3\pi\frac{\rho}{k^3}-1}.\label{eq:Ioffe}
\end{equation}
A more quantitative analysis is presented in Fig.~\ref{Fig2}, where the proportion of each kind of modes is shown, for different densities and for a fixed particle number. We first note that the proportion of superradiant modes decreases as for a fixed number of atoms, an increasing density comes along a reduced volume. This is consistent with the fact that in the subwavelength regime, a single superradiant value has a rate that tends to $N$, all other modes being subradiant.

More surprisingly, the remaining long-lived modes are largely dominated by subradiant ones, which typically constitute $\sim75\%$ of the $N$ modes. Indeed, even for a density $\rho=10\rho_c$, localized modes represent at most one-fifth of all modes. The localization scenario where the transport properties of the system operate a transition from a conductor to an insulator, is clearly not valid if the localized modes cannot be addressed specifically. Furthermore, working at fixed particle number implies that the system size changes significantly when changing the density. Increasing the particle number from $N=1500$ to $N=7000$ leads to a slight increase of the proportion of superradiant modes (as the optical thickness increases), the proportion of subradiant modes decreasing correspondingly, and that of localized modes being almost unaltered. Hence, finite-size effects do not appear to be responsible for the low proportion of localized modes. This clearly questions the notion of the localization transition in that system, especially considering the existing debate on the possible signatures of the transition. For example, no exponential decay of the transmission with the system size, a paradigmatic signature of the localization regime, has been reported up to date in 3D systems. The existence of a large population of other modes (superradiant and subradiant), that could support the transport of light through the sample, may contribute to the lack of clear signatures of transition, despite the statistical eigenvalue analysis that favors long lifetimes modes predicts a transition~\cite{SkipetrovAbsence}. Let us comment that a partial localization was found by Stockman et al. in the context of surface plasmons~\cite{partial_localize}, which was attributed to long-range dipolar interaction.

\begin{figure}
   \centering
    \includegraphics[width=.9\columnwidth]{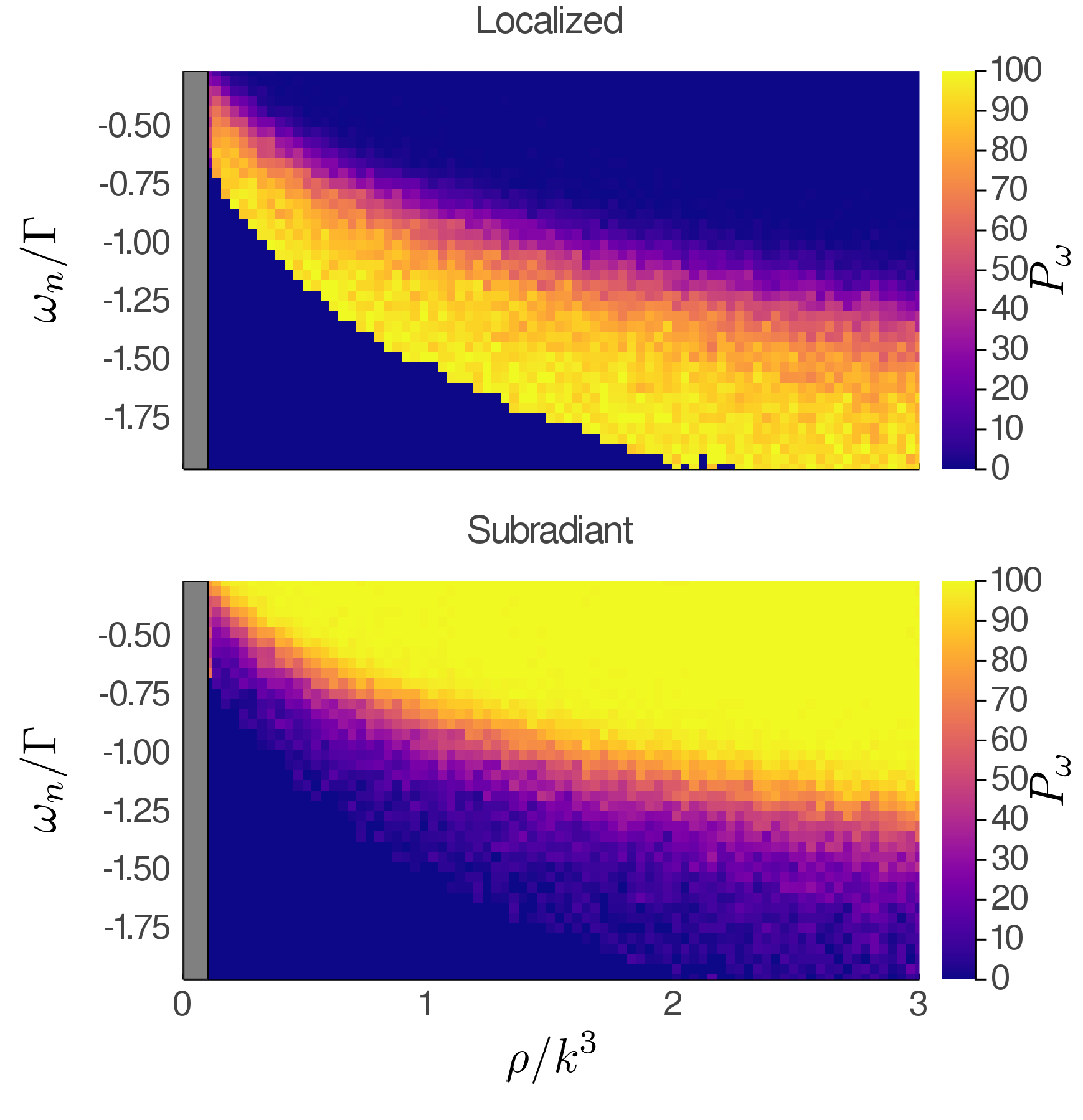}
    \caption{Relative proportion $P_\omega$ of localized and subradiant modes for a narrow ($\delta\omega=0.021$) range of energy, as a function of the system density. Simulations realized with $N=1300$ particles. A threshold of 15 was imposed on the number of modes in that energy range to plot the resulting proportions. The proportion of superradiant modes is not represented, as their number is usually below the threshold.}
    \label{Fig3_0}
\end{figure}

A refined analysis on the distribution of energy of these modes reveals that localized modes are concentrated over specific ranges of energy, bounded by the mobility edges~\cite{Skipetrov2018}. As can be observed in Fig.\ref{Fig3_0}, almost all modes are localized in the proper range of negative detuning, whereas in a different range of energy around the resonance, almost all modes are subradiant. Superradiant modes represent always a minority of modes, yet as we shall now see, their relative contribution cannot be neglected.

Localized modes represent a minority of all modes, but its quasi-totality for specific energy ranges. An open question is how an incoming wave will couple to them. This coupling can be evaluated by studying the projection of the pump onto the scattering modes. The solution to Eq.~\eqref{eq2} decomposes as:
\begin{equation}
    \vec\beta(t) = \sum_n \alpha_n \Psi_n e^{\lambda_n t},
    \label{eq_beta_diag}
\end{equation}
where $\alpha_n$ represents the projection coefficient. Thus, the set of $|\alpha_n|^2$ represents the population of the modes~\cite{population}. We here focus on the steady-state regime, for which the $\alpha_n$s are obtained from Eq.~\eqref{eq2} as:
\begin{equation}
    \label{eq_beta_esta}
    \vec{\alpha} = \frac{i}{2}\Psi^{-1}M^{-1}\vec{\Omega}
\end{equation}


In order to understand if it is possible to address the localized modes specifically, we study the mode population defined as 
\begin{equation}
  P_\mathrm{Loc}=\dfrac{\sum\limits_{n\in \{\mathrm{Loc}\}}\left|\alpha_n\right|^2}{\sum\limits_{n=1}^N \left|\alpha_n\right|^2},
\end{equation}
where $\{\mathrm{Loc}\}$ refers to the ensembles of localized modes ($P_\mathrm{SR}$ and $P_\mathrm{Sub}$ are defined in a similar way).

\begin{figure}
   \centering
    \includegraphics[width=\columnwidth]{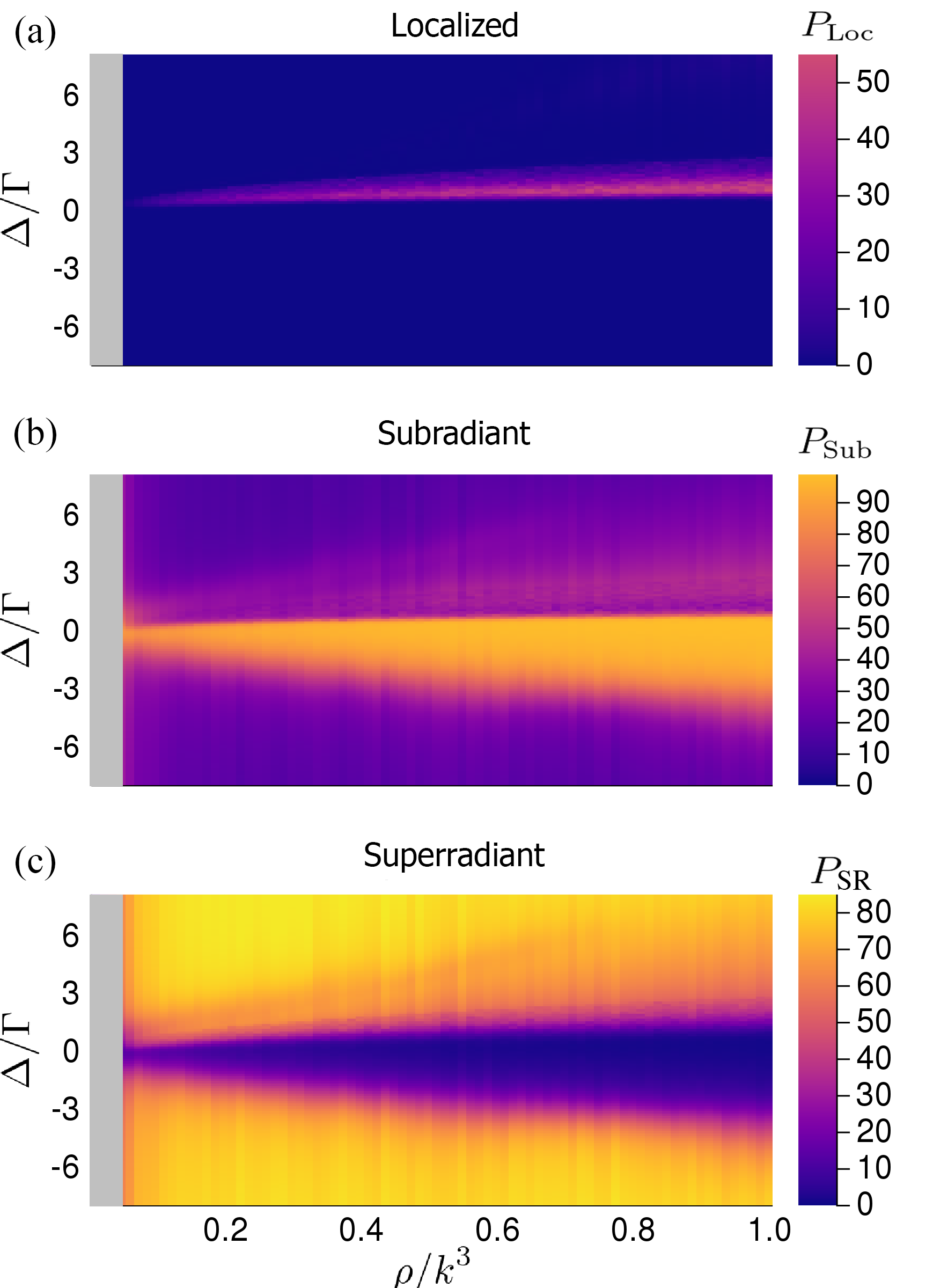}
    \caption{Population in the localized, subradiant and superradiant modes to a plane-wave, as a function of density and detuning. Simulations realized with a particle number $N=500$, with an averaging over $100$ realizations for each value of detuning and density. The gray area for low densities correspond to a regime of low resonant optical thickness, where superradiance and subradiance are not well formed; this area reduces as the number of particles is increased.}
    \label{Fig3}
\end{figure}

In order to address specifically the localized modes, the detuning $\Delta$ is tuned over the resonance, in addition to the atomic density. As can be observed in Figure~\ref{Fig3}(a), the localized modes are addressed more specifically for a detuning of $\Delta\sim\Gamma$, yet even then, and for densities well above the critical one, at most half of the incoming plane-wave couples to localized modes. The fact that at least half of the incoming wave projects on superradiant and subradiant modes may explain why a clear signature of the phase transition is difficult to identify in steady-state transmission~\cite{Skipetrov2016}. Indeed, these modes are unrelated to localization, as they are already present in the dilute regime, well below the critical density~\cite{Guerin2016,Araujo2016,Roof2016}.

We also see that subradiant modes are addressed with a larger population for moderate but negative detuning (see Fig.\ref{Fig3}(b)), where almost all the plane wave may project on these. At large absolute values of the detuning, when the cloud turns transparent, superradiant modes are addressed predominantly. This is consistent with the picture of atoms illuminated homogeneously, with a phase given by that of the laser, as exemplified by the superradiant timed Dicke state~\cite{Scully2006,Cottier2018}. Note that we have checked that using a Gaussian beam with a waist smaller than the cloud radius (for a ratio as low as one third) does not alter significantly the population of localized modes.



\begin{figure}
   \centering
    \includegraphics[width=\columnwidth]{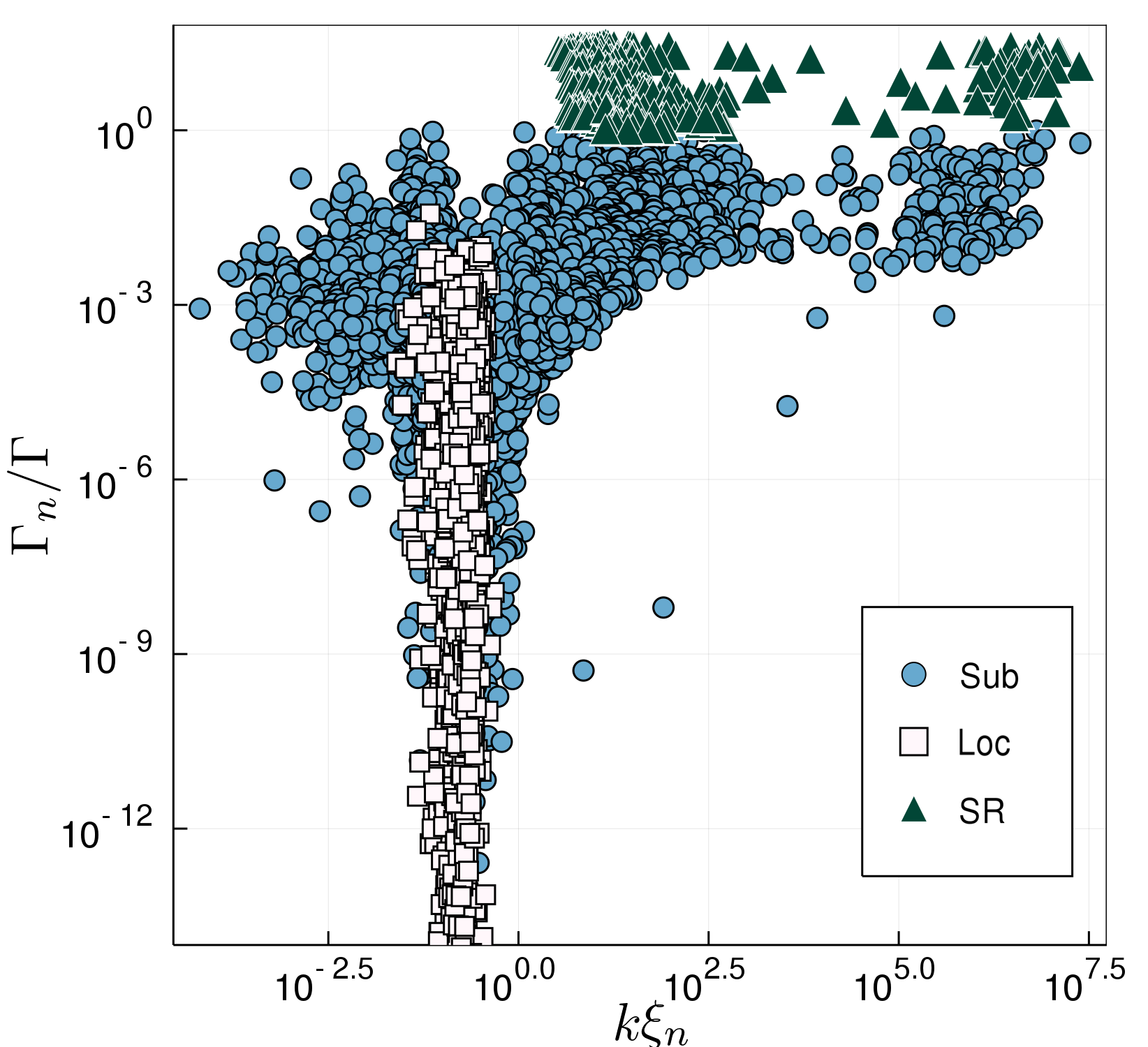}
    \caption{(Absence of) correlations between the decay rate and the localization length of the localized modes. Simulation realized for a cloud of density $\rho=k^3$, with $N=9000$ particles.}
    \label{Fig4}
\end{figure}

We now turn to the lifetimes (widths) of the eigenmodes. As can be seen in Fig.\ref{Fig4}, the localized modes present a wide range of lifetimes, that extends over many orders of magnitude. Nevertheless, as it has been observed in the 2D case~\cite{Maximo}, there is no clear correlation between the lifetime and localization length of the modes.

\begin{figure}
   \centering
    \includegraphics[width=\columnwidth]{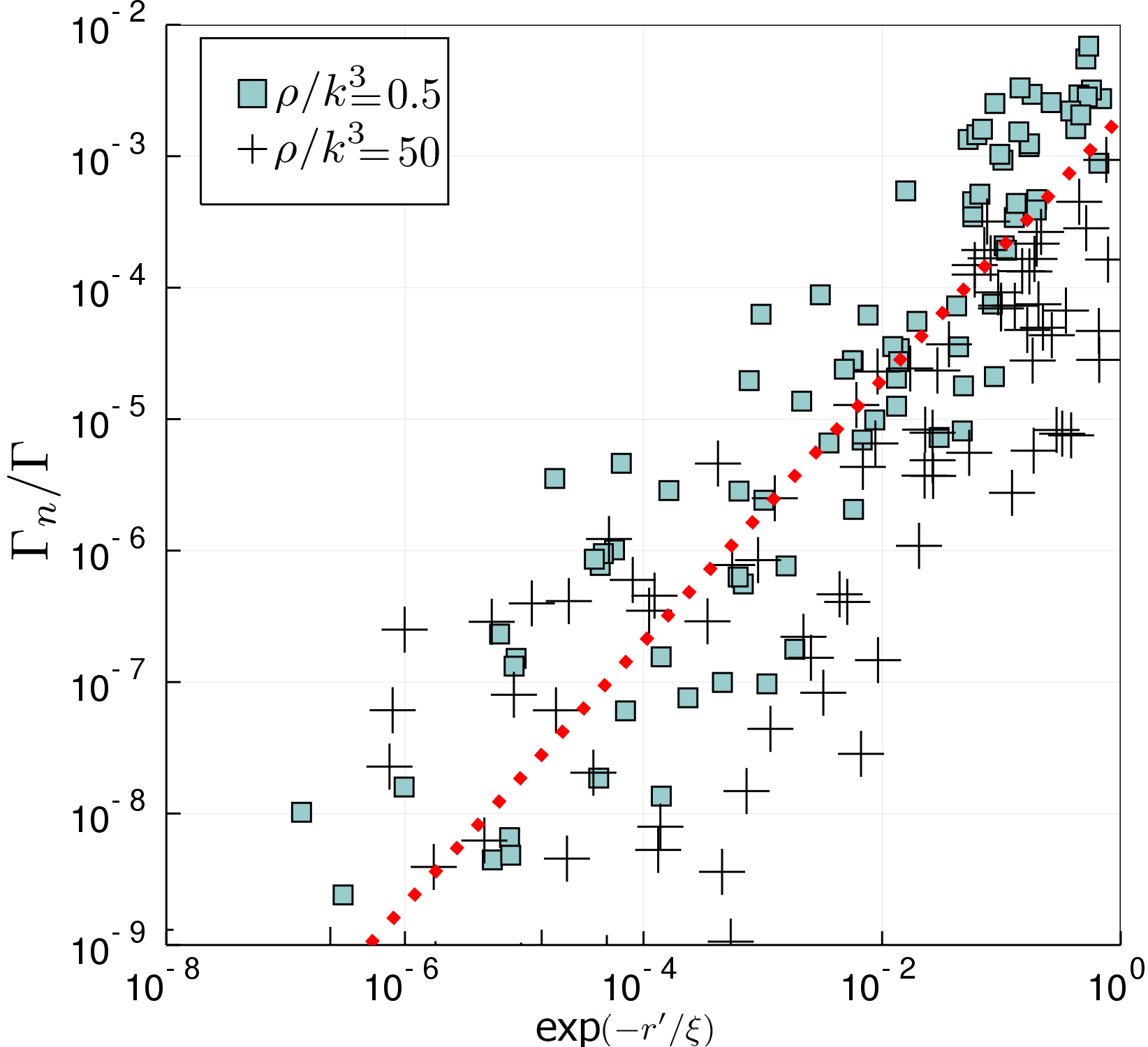}
    \caption{Decay rate of the modes as a function of the ratio between their distance to the boundary $r'=(R-|\mathbf{r}_{\mathbf{CM}}^n|)$ and their localization length $\xi$. The dotted line corresponds to $\Gamma_n=C\exp(-r'/\xi_n)$, with $C=2.10^{-3}$. The simulation for density $\rho=0.5k^3$, resp. $50k^3$, presents localization lengths of order $\xi \sim 0.4/k$, resp. $\sim 0.07/k$. Simulations realized with $N=3000$ particles.}
    \label{Fig5}
\end{figure}

The main source of the observed broadening in the modes linewidth can be found in finite-size effects: The proximity to the system boundary results in a leakage of the mode, which appears to be well described by $\Gamma_\textrm{leak}=C\exp(-r'/\xi)$~\cite{Pinheiro2004}, where $r'=R-|\mathbf{r}_{ \mathbf{CM} }|$ is the distance of the mode center-of-mass to the boundary, and $C\approx2.10^{-3}$. This effect is illustrated in Fig.\ref{Fig5}, where a clear correlation between the distance to the system's boundary and the mode lifetime is found, for two very different sets of parameters (corresponding to $\xi\sim 0.4/k$ and $\xi\sim 0.07/k$). In the regime we have explored, the decay rates $\Gamma_n$ appear to be dominated by this effect, and the Heisenberg and Thouless times do not appear to affect these modes~\cite{Akkermans2007}. The scaling law for $\Gamma_\textrm{leak}$ was observed for decay rates as small as the numerical precision of our simulations allows ($\sim 10^{-14}$). This suggests that for most experimentaly accessible parameters, the lifetime of localized modes is determined by their distance to the boundary and that possible size-independent limits on the lifetime of the localized modes only appear at much longer times.

\section{Conclusions \& Perspectives}\label{Sec:Conclusion}
Across this work, we have studied the population of modes throughout the Anderson localization phase transition for 3D light scattering. Spatially-extended subradiant and superradiant modes are present over a broad range of energies, and represent the majority of modes. Although between the mobility edges, almost all modes are localized, tuning the pump frequency close to the resonant energy of the localized modes does not allow one to couple more than half of the incoming wave to these localized modes, at least for the systems accessible numerically. Hence, the contribution from other modes to the transport of light cannot be excluded in this regime.

A natural question is whether the transport of light can really be put to a halt thanks to disorder, with the paradigmic exponential decay of diffuse transmission with the system length~\cite{Berry1997}.  On the contrary, so far, only an increase of the transport at higher densities, as compared to the diffusion prediction, has been reported~\cite{Guerin2016b}. The absence of analytical tools and the strong limitations of numerical simulations for these systems definitely represent an obstacle to understand the specificity of 3D light scattering. In this regard, the correlation between the lifetime of localized modes and the system size observed in our  work is are a good illustration of the importance of finite-size effects, and of the subsequent limitations in simulating small systems. As for the steady-state transport, a dedicated study will be required to understand how to couple specifically to the localized modes, since one has to consider not only the gap width (given by the mobility edges), but also the linewidth of the neighbouring superradiant and subradiant modes.

\acknowledgments
Part of this work was performed in the framework of the European Training network ColOpt, which is funded by the European Union Horizon 2020 program under the Marie Skodowska-Curie action, grant agreement 72146. The work was also supported by the ANR (project LOVE, Grant No. ANR-14-CE26-0032). R. B. benefited from Grants from São Paulo Research Foundation (FAPESP) (Grants Nos. 2014/01491-0, 2015/50422-4 and 2018/01447-2) and from the National Council for Scientific and Technological Development (CNPq) Grant Nos. 302981/2017-9 and 409946/2018-4. R. B. and R. K. received support from project CAPES-COFECUB (Ph879-17/CAPES 88887.130197/2017-01). The Titan X Pascal used for this research was donated by the NVIDIA Corporation.

%

\bibliography{main}
\bibliographystyle{unsrt}
\end{document}